\documentclass[preprint, authoryear]{elsarticle}
\usepackage{soul, color}
\usepackage{colortbl}
\usepackage{bm}
\usepackage{graphicx}
\usepackage{booktabs}
\usepackage{caption}
\usepackage{arydshln}
\usepackage{subcaption}
\usepackage{amsmath}
\usepackage{amssymb}
\usepackage{amsthm}
\usepackage{dsfont}
\usepackage{bbold}
\usepackage{listings}
\usepackage{setspace}
\usepackage{url}
\usepackage{verbatim}
\usepackage{dsfont}
\usepackage{float}
\usepackage{xspace}
\usepackage{multirow}
\usepackage{makecell}
\usepackage[shortlabels]{enumitem}
\usepackage{tikz}
\usetikzlibrary{shapes,positioning,arrows,calc,trees, matrix, arrows.meta}

\tikzset{
  mynode/.style={circle, draw, minimum size=1cm}
}

\tikzset{
  treenode/.style={
    rectangle, draw, align=center, rounded corners=4pt, font=\large,
    minimum width=1cm, minimum height=0.8cm
  }
}

\newcommand{\pihat}{\widehat{\pi}}

\newcommand{\Abf}{\mathbf{A}}

\newcommand{\bbf}{\mathbf{b}}
\newcommand{\ubf}{\mathbf{u}}

\newcommand{\rr}{\mathbb{R}}
\newcommand{\nn}{\mathbb{N}}

\newcommand{\NB}{\mathrm{NB}}

\newcommand{\met}{e2eTD\xspace}

\definecolor{codegreen}{rgb}{0,0.6,0}
\definecolor{codeorange}{RGB}{250,100,0}
\definecolor{codepurple}{rgb}{0.58,0,0.82}
\definecolor{codeblue}{RGB}{50,50,250}

\definecolor{codegray}{rgb}{0.5,0.5,0.5}
\definecolor{codepurple}{rgb}{0.58,0,0.82}
\definecolor{backcolour}{rgb}{0.95,0.95,0.92}
\definecolor{Gray}{gray}{0.8}
\definecolor{lightyellow}{RGB}{245,238,197}
\newcolumntype{a}{>{\columncolor{lightyellow}}c}
\newcolumntype{b}{>{\columncolor{backcolour}}c}

\lstdefinestyle{mystyle}{
    backgroundcolor=\color{backcolour},
    commentstyle=\color{codegreen},
    keywordstyle=\color{magenta},
    numberstyle=\tiny\color{codegray},
    stringstyle=\color{codepurple},
    basicstyle=\ttfamily\footnotesize,
    breakatwhitespace=false,
    breaklines=true,
    captionpos=b,
    keepspaces=true,
    numbers=left,
    numbersep=5pt,
    showspaces=false,
    showstringspaces=false,
    showtabs=false,
    tabsize=2
}

\DeclareFontFamily{U}{mathx}{}
\DeclareFontShape{U}{mathx}{m}{n}{<-> mathx10}{}
\DeclareSymbolFont{mathx}{U}{mathx}{m}{n}
\DeclareMathAccent{\widehat}{0}{mathx}{"70}
\DeclareMathAccent{\widecheck}{0}{mathx}{"71}

\newtheorem{remark}{Remark}

\begin{document}
\begin{frontmatter}
\title{End-to-end probabilistic hierarchical forecasting of large hierarchies via probabilistic top-down}

\author[1]{Lorenzo Zambon\corref{cor1}}
\ead{lorenzo.zambon@supsi.ch}

\author[1]{Dario Azzimonti}
\ead{dario.azzimonti@supsi.ch}

\author[1]{Giorgio Corani}
\ead{giorgio.corani@supsi.ch}

\cortext[cor1]{Corresponding author}

\affiliation[1]{organization={SUPSI, Istituto Dalle Molle di Studi sull'Intelligenza Artificiale (IDSIA)},
city={Lugano},
country={Switzerland}}

\begin{abstract}
Retail and supply chain operations rely on demand forecasts to drive decisions, from replenishment at the product level to capacity planning at the store level.
These forecasts should be probabilistic, to allow risk-aware decisions, and coherent across the aggregation hierarchy, so that decisions taken at different levels 
are not based on conflicting demand forecasts.
Producing such forecasts is computationally demanding; at retail scale, with hierarchies of hundreds of thousands of time series, this cost becomes a primary concern.
Existing two-step forecast-then-reconcile procedures and end-to-end neural models scale poorly, rely on restrictive assumptions, or require specialized hardware and engineering effort.
We propose e2eTD, a scalable method for coherent probabilistic forecasting of large hierarchical and grouped time series. 
e2eTD directly forecasts only a small subset of aggregate series (about 0.3\% of the hierarchy in our experiments), which are smoother and thus more predictable than the intermittent bottom series.
The resulting forecast samples are propagated to the bottom level through a novel probabilistic top-down sampling algorithm, in which the historical disaggregation proportions are modeled as joint distributions, estimated in-sample.
Since the bottom samples retain the cross-series dependence, summing them yields coherent forecasts for all aggregation levels.
On the M5 and Favorita datasets, e2eTD achieves the lowest weighted scaled pinball loss across aggregation levels among all competing methods; it would have ranked 11th of 892 teams in the M5 Uncertainty competition.
On a standard laptop, e2eTD runs in about five minutes on M5 ($\sim$40K series) and twenty minutes on Favorita ($\sim$300K series).
\end{abstract}

\end{frontmatter}

\section{Introduction}
\label{sec: intro}

In retail and supply chain logistics, demand data are naturally organized into hierarchical or grouped structures \citep{fildes2022retail, babai2022demand}:
individual stock keeping units (SKUs) are aggregated along several dimensions, such as store, region, and product category \citep{makridakis2022m5}. 
Accurate forecasts are needed at every level to support effective planning and control: for example, replenishment at the SKU level, capacity and assortment at the store level, and strategic planning at the regional or national level \citep{silver2016inventory, babai2022demand}.

An important requirement in hierarchical settings is forecast coherency. 
Forecasts are coherent when they satisfy the hierarchical constraints across all levels of aggregation:
for example, the forecast for total sales must be equal to the sum of the forecasts for each constituent item or store. 
This property is essential for operational alignment, as incoherent forecasts can lead to misaligned decisions across different business units \citep{babai2022demand}.
Forecasts should also be probabilistic, providing full predictive distributions rather than single point forecasts \citep{kolassa2016evaluating}, since quantifying uncertainty enables the risk-aware decision-making necessary for robust planning \citep{prak2019general, spiliotis2021product}.
In practice, a probabilistic forecast is often represented as a collection of samples, each characterizing a possible future scenario.
In hierarchical settings, each scenario must be consistent across all levels:
probabilistic coherency requires that each sample from the joint predictive distribution satisfies the aggregation constraints \citep{panagiotelis2023probabilistic}.

Computational cost is a further concern at retail scale.
A single large retailer may need to refresh on the order of 
$10^9$ store~$\times$~SKU forecasts on a daily or weekly basis 
\citep{seaman2018considerations};
\citet{petropoulos2025wielding} estimate that at Walmart's online scale, switching to a fast-and-frugal model can save tens of millions of dollars per year. 
Practitioners explicitly acknowledge this trade-off: \citet{yelland2019forecasting} report that Target deliberately accepts small accuracy losses to obtain computational savings.
These savings extend to the environmental footprint of large-scale forecasting \citep{petropoulos2025wielding};
\citet{schwartz2020green} argue that efficiency should be considered as an evaluation criterion alongside accuracy.

Satisfying these requirements is further complicated by the nature of retail hierarchies, which are typically characterized by the prevalence of intermittent time series at the bottom level: individual item-store combinations produce long runs of zero observations interspersed with sporadic spikes, and exhibit a low signal-to-noise ratio that makes them hard to forecast accurately. 
Aggregate series, by contrast, are smoother and more predictable, benefiting from the averaging effect of multiple contributing components \citep{oliveira2019assessing}.

Existing approaches to hierarchical forecasting fall into two main families. Two-step \textit{forecast-then-reconcile} procedures produce base forecasts independently for each series and then adjust them to satisfy the aggregation constraints \citep{athanasopoulos2024forecast}. 
Computing base forecasts for every series is expensive at scale, and the Gaussian assumption underlying scalable reconciliation is untenable for intermittent counts.
\textit{End-to-end} methods instead learn a single model that produces coherent forecasts jointly across the hierarchy; they typically rely on neural architectures that require substantial computational resources and model engineering \citep{rangapuram2021end, olivares2024probabilistic, das23dirichlet}. 
A separate line of work addresses intermittent series directly, based on neural networks \citep{kourentzes2013intermittent}, state-space models \citep{svetunkov2023}, or more scalable alternatives \citep{long2025};
these methods, however, 
do not produce coherent forecasts.
An effective and scalable method for coherent probabilistic forecasting of large hierarchies with intermittent bottom series is therefore still missing.

We fill this gap by proposing \met: 
it directly forecasts only a small subset of the smooth upper series (about 0.3\% of the total in our experiments); a novel probabilistic top-down algorithm then propagates the forecast samples to the bottom level, using in-sample joint distributions as historical proportions. Summing the resulting joint bottom-level samples yields coherent forecasts across all levels. As a result, \met scales to very large hierarchies: on M5 ($\approx$30k bottom and $\approx$10k upper series, \citealp{makridakis2022m5}), our implementation runs in under five minutes on a standard laptop, with no specialized hardware.

The rest of the paper is organized as follows. 
Sect.~\ref{sec:prob-forecasting} sets up the problem of coherent probabilistic hierarchical forecasting, and Sect.~\ref{sec:literature_prob_hier} reviews related work. Sect.~\ref{sec:e2etd} introduces the e2eTD method and explains the probabilistic top-down sampling algorithm and the computational strategies that make it scalable.
Sect.~\ref{sec:emp_evaluation} presents the empirical evaluation on the M5 and Favorita datasets, including accuracy, computational cost, and an ablation study. 
Sect.~\ref{sec: conclusions} discusses the results and outlines directions for future work.

\section{Probabilistic hierarchical forecasting}
\label{sec:background}

\begin{figure}[!h]
\vspace{2 mm}
    \begin{center}
        \begin{tikzpicture}[every node/.style={rectangle,draw,align=center,rounded corners=4,font=\large},level distance=1.cm,
  level 1/.style={sibling distance=5.cm, edge from parent fork down},
  level 2/.style={sibling distance=2.1cm, edge from parent fork down},
  level 3/.style={sibling distance=1.2cm, edge from parent fork down}
]
  \node[fill=black!20,minimum width=1cm] {$Z$}
    child {node[fill=black!10,minimum width=1cm] {$A$}
      child {node[fill=black!0,minimum width=1cm] {$AA$}}
      child {node[fill=black!0,minimum width=1cm] {$AB$}}
    }
    child {node[fill=black!10,minimum width=1cm] {$B$}
      child {node[fill=black!0,minimum width=1cm] {$BA$}}
      child {node[fill=black!0,minimum width=1cm] {$BB$}}
    };
\end{tikzpicture}
    \caption{A hierarchy with $4$ bottom and $3$ upper time series.}
    \label{fig:simple_hier}
    \end{center}
\end{figure}
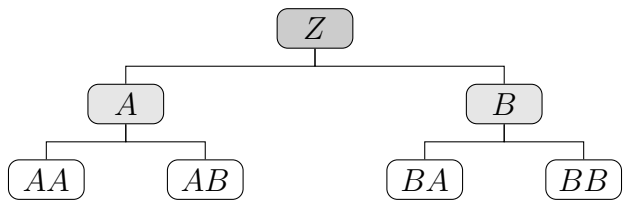

\paragraph{Hierarchical forecasting}

Hierarchical time series are collections of time series organized across multiple levels of aggregation, where higher levels represent aggregated totals and lower levels correspond to finer disaggregations.
%
The most disaggregated series are called \textit{bottom}, while the others are called \textit{upper}:
in the hierarchy of Fig.~\ref{fig:simple_hier},  
$AA, AB, BA, BB$ are the bottom series and $A, B, Z$ are the upper series. 
At each time $t$, the observed values satisfy
\begin{equation*}
A_t = AA_t + AB_t, \quad B_t = BA_t + BB_t, \quad Z_t = A_t + B_t = AA_t + AB_t + BA_t + BB_t.
\end{equation*}
More generally, let $\bbf_t\in\rr^{n_b}$ denote the vector of bottom time series at time $t$ and $\ubf_t\in\rr^{n_u}$ the vector of upper series. 
The hierarchical constraints between series can be written compactly as
\begin{equation}
\ubf_t=\Abf\,\bbf_t,
\label{eq:aggregation-matrix}
\end{equation}
where $\Abf\in\rr^{n_u\times n_b}$ is the \textit{aggregation matrix}, which is made of 0s and 1s and specifies how the bottom series aggregate to the upper series.
Often, time series are not strictly hierarchical but \textit{grouped}: the structure cannot be expressed as a single tree, as the series are aggregated across different directions; in the following, we use the term hierarchical to cover both cases.
Hierarchical forecasts should be \textit{coherent}, i.e., they should satisfy the same constraints as the observations
\citep{athanasopoulos2024forecast}.



\paragraph{Probabilistic forecasting}
\label{sec:prob-forecasting}

Probabilistic forecasting quantifies the uncertainty of the prediction by returning a distribution rather than a point estimate.
Probabilistic forecasts may be specified as parametric predictive distributions (e.g., Poisson) or represented by samples; in both cases, point, interval, and quantile summaries can be derived. 
In this paper, we adopt a sample-based representation, where each sample corresponds to a possible future scenario.
For hierarchical time series, each scenario $i$ is a multivariate draw that includes the future values of all series in the hierarchy:
\[
\big[\ubf_{T+h\,|\,T}^{(i)},\;\bbf_{T+h\,|\,T}^{(i)}\big],
\]
representing one possible realization of the joint predictive distribution.
Forecasts are computed at some time $T$ for a future time $T+h$; for ease of notation, in the following we drop the time subscripts.
Probabilistic coherence \citep{panagiotelis2023probabilistic} requires that every draw satisfies the hierarchical constraints:
\[
\ubf^{(i)} = \Abf \bbf^{(i)}, \quad \text{for all } \; i,
\]
so that each simulated future scenario is consistent across all levels of the hierarchy.

\paragraph{The value of coherent probabilistic forecasts}
In production and inventory settings, point forecasts are often not sufficient \citep{fildes2022retail}.
Replenishment, for instance, trades off holding costs against stock-out costs, so the cost-optimal order quantity is a quantile of the predictive distribution rather than its mean \citep{silver2016inventory}.
In retail this quantile is typically high, since it matches the targeted service level \citep{boylan2006accuracy, spiliotis2021product}, and when safety stocks must cover multi-period lead times, the full distribution is needed to obtain quantiles of cumulative future demand \citep{kolassa2016evaluating}.
The need is particularly important at the SKU level, which drives most replenishment decisions and where demand is typically intermittent: 
the mean is typically fractional and thus not a usable order quantity, while the median is often zero and hence useless for replenishment \citep{syntetos2005accuracy, kolassa2016evaluating}.
Probabilistic forecasts instead support decisions based directly on the relevant quantile.

In a production setting, forecasts are needed at multiple aggregation levels: replenishment is planned at the store~$\times$~SKU level, capacity and assortment at the store level, and strategic decisions at the category or national level;
these decisions are coordinated to align operational, tactical, and strategic plans across the hierarchy
\citep{babai2022demand, kremer2016sum}. 
Forecast coherence guarantees that decisions made at different levels rely on the same view of demand \citep{kourentzes2019cross, pritularga2021stochastic}, preventing cross-level inconsistencies that can lead to misaligned plans \citep{athanasopoulos2024forecast}.
For probabilistic forecasts this requirement goes beyond the coherence of the forecast means: the entire joint distribution across series must satisfy the aggregation constraints \citep{panagiotelis2023probabilistic}, so that risk is assessed consistently across levels \citep{taieb2021hierarchical}.


\section{Related work}
\label{sec:literature_prob_hier}

We briefly review the main approaches to probabilistic hierarchical forecasting. 
We focus in particular on methods that produce coherent forecasts and discuss the challenges that arise in large hierarchies with low-count bottom time series.

\paragraph{Single-level approaches}
The earliest hierarchical forecasting methods compute forecasts at a single level of aggregation and then propagate them to the rest of the hierarchy \citep{Athanasopoulos2009tourism}.
The \textit{bottom-up} approach forecasts each bottom series and aggregates the forecasts upward;
\textit{top-down} forecasts only the top level and disaggregates using historical proportions; \textit{middle-out} forecasts a single intermediate level and propagates the result both upward and downward.
Relying on a single aggregation level can yield information loss and introduce modelling and estimation risks \citep{kourentzes2017demand, athanasopoulos2024forecast}; bottom-up is particularly affected, as bottom series in retail often have a low signal-to-noise ratio \citep{oliveira2019assessing}.
%

Bottom-up can be extended to the probabilistic setting by summing samples rather than point forecasts, but this requires modeling cross-series dependence: assuming independence typically underestimates aggregate uncertainty \citep{zambon2024probabilistic}, 
whereas existing empirical-copula approaches \citep{taieb2021hierarchical} are tailored to specific applications and do not readily scale to large hierarchies with short training sets \citep{panagiotelis2023probabilistic}.
Top-down and middle-out approaches are attractive for large hierarchies, as they forecast only a small number of upper series. Extending them to the probabilistic setting, however, is not straightforward, as it requires modeling the distribution of disaggregation proportions rather than applying deterministic ratios.
\citet{das23dirichlet} learn these proportions through a shared global Dirichlet neural network.
The method requires training a deep model and its scalability to large hierarchies has not been demonstrated, as it is evaluated on reduced versions of the M5 and Favorita datasets. 
\citet{long2025} instead disaggregate only the point forecasts, computed at an intermediate level, using historical proportions; bottom-level forecasts are assumed to follow a negative binomial distribution, with variances estimated in-sample.
This approach is computationally efficient and generally accurate at the bottom level, but it does not yield coherent forecasts for the full hierarchy: only marginal bottom-level distributions are produced, with no model of cross-series dependence, so upper-level predictive distributions cannot be recovered.

\paragraph{Reconciliation-based approaches}
A large body of hierarchical forecasting literature addresses coherency via a two-step procedure. 
First, base forecasts are independently produced for all the time series of the hierarchy;
then, forecasts are adjusted to satisfy the aggregation constraints (\textit{reconciliation}).
Early work on forecast reconciliation only focused on point forecasts \citep{hyndman2011optimal, Wickramasuriya2019}.
More recently, this has been extended to \textit{probabilistic reconciliation}, which provides reconciled predictive distributions.
If the base forecasts are assumed to be jointly Gaussian,
the reconciled distribution admits a closed-form solution \citep{wickramasuriya2023probabilistic, zambon2024properties};
however, assuming normality is untenable for intermittent count series \citep{spiliotis2021product}.
In the non-Gaussian case, probabilistic coherence is usually achieved by reconciling each sample from the base forecast distribution via a projection map \citep{panagiotelis2023probabilistic, girolimetto2024point}.
However, this approach is limited to continuous forecasts, may violate non-negativity,
and scales poorly in large hierarchies as it relies on either high-dimensional matrix inversions or numerical optimization.
An alternative approach is based on conditioning:
it naturally handles non-negative discrete data,
and also mixed-type hierarchies with low-count bottom series and smooth aggregated series \citep{zambon2024efficient, zambon2024probabilistic}.
Yet, it relies on sampling algorithms that require strong distributional assumptions and can be computationally expensive in high dimensions.
In short, while forecast-then-reconcile paradigms guarantee coherence, their scalability is often limited by the computation of the base forecasts for all the time series in the hierarchy and the significant overhead of the reconciliation process.

\paragraph{End-to-end models} 
In contrast to two-step procedures, end-to-end methods generate coherent probabilistic forecasts within a unified framework. 
These approaches learn the joint distribution of the hierarchy globally, capturing cross-series dependencies and integrating the hierarchical constraints into the model. 
\cite{rangapuram2021end} incorporate a differentiable analytical reconciliation step directly into the optimization objective. 
However, this technique relies on the reparameterization trick, restricting the method to continuous distributions.
\cite{kamarthi2024large} enforce coherence via a penalty term in the loss function, which acts as a soft regularization rather than imposing strict coherence. The method selects Gaussian or Poisson forecast distributions based on a sparsity criterion, relying on independence assumptions and approximations for mixed aggregation levels that limit its statistical soundness.
\citet{olivares2024probabilistic} propose the Deep Poisson Mixture Network, which provides a joint predictive distribution that is coherent by construction. 
\citet{olivares2023hint} extend this line of work with HINT, a modular framework that augments neural forecasting architectures  with a hierarchical multivariate mixture output and enforces coherence via bootstrap sample reconciliation. 
Both methods, however, model the joint distribution of the hierarchy through a single  mixture model shared across the hierarchy: dependence between series is captured implicitly via shared latent variables, and all series are constrained to follow the same mixture family. This can be limiting in retail, where bottom series are intermittent counts while upper aggregates are smooth.
A common feature of all these end-to-end models is their reliance on deep neural networks, which can be computationally intensive and typically require significant engineering effort and GPU hardware for training.

\paragraph{Static distributional baselines}

Other approaches bypass modeling time-series dynamics entirely, producing probabilistic forecasts directly from in-sample observations at very low computational cost. 
This is particularly attractive for intermittent bottom series,
which constitute the majority of the hierarchy and for which conventional time series models often struggle.
\citet{kolassa2016evaluating} advocates fitting static parametric distributions (such as Poisson or negative binomial) to each series, or simply using in-sample empirical quantiles. \citet{spiliotis2021product} show on the $\approx$30K bottom series of the M5 dataset that these approaches achieve accuracy comparable to, or better than, statistical and machine learning models, at a fraction of the computational cost.
However, these methods are not designed for coherence: per-series parametric fits are generally incoherent across the hierarchy.
The empirical baseline is an exception: by treating historical joint observations as samples from the joint distribution, the resulting forecast is coherent by construction.
A variant uses seasonal empirical quantiles, fitting a separate distribution for each position within the seasonal cycle (e.g., the same weekday); it can capture weekly patterns typical of retail data at the price of fewer observations per fit. 
All these approaches rely on time-invariant distributions and therefore fail to account for trend, level shifts, or external drivers; their accuracy degrades especially at upper aggregation levels, where such dynamics are more pronounced.

\section{End-to-end top-down}
\label{sec:e2etd}
 
\begin{figure}[h!]
  \centering
  \includegraphics[width=1.\textwidth]{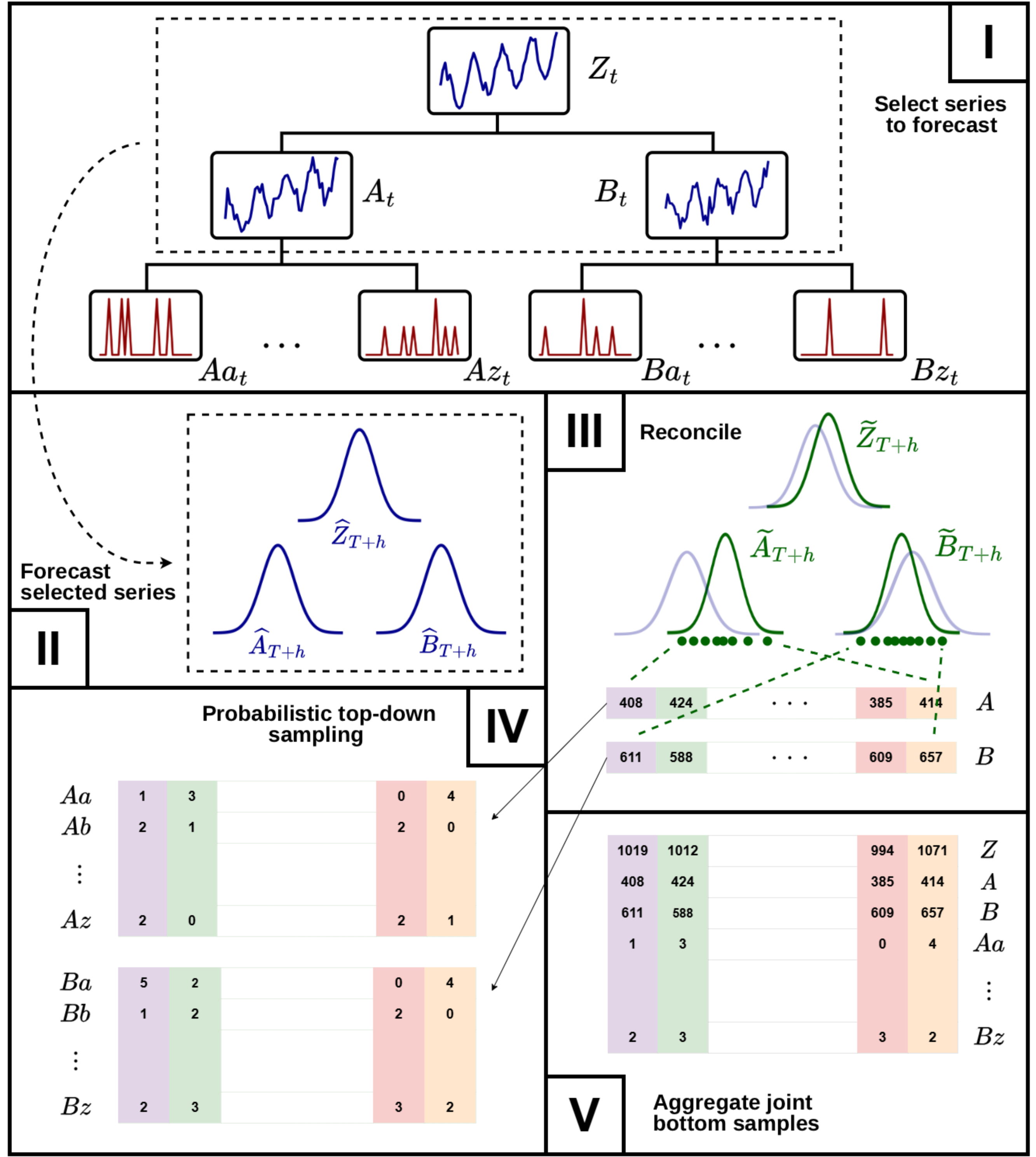}
  \caption{
  Overview of the proposed \met methodology. (I) A subhierarchy of smooth, predictable upper time series is selected. (II) Probabilistic forecasts are generated for the selected subhierarchy. (III) Forecasts are reconciled to ensure coherence across the upper levels; samples are drawn from the reconciled multivariate distribution for the subhierarchy's lowest level. (IV) A probabilistic top-down sampling algorithm disaggregates these samples to generate joint samples for the bottom series. (V) Final coherent joint forecasts for the entire hierarchy are computed by bottom-up aggregation.
  }
  \label{fig:e2etd}
\end{figure}

Our method, which we call \textit{\met},
produces coherent probabilistic forecasts for the entire hierarchy in the form of \textit{joint samples}.
The main steps of \met are detailed below and represented in Fig.~\ref{fig:e2etd}.

\paragraph{I. Select time series to forecast}
We select a subset of time series to forecast among the most aggregate series, as they often contain more signal.
The subset must form a subhierarchy covering all bottom series; i.e., every bottom series is a descendant of at least one selected series.

\paragraph{II. Forecast selected series}
Our approach can employ any probabilistic forecasting method.
In our experiments, we use 
exponential smoothing (ETS) with automatic model selection \citep[Ch.~8]{hyndman2021fpp3},
which we fit independently for each selected series.

\paragraph{III. Reconcile}
We combine the forecasts of the selected upper time series by applying Gaussian probabilistic reconciliation \citep{wickramasuriya2023probabilistic, zambon2024properties}.
We draw joint samples from the reconciled multivariate normal distribution at the
lowest level of the subhierarchy.
Samples are rounded and clipped to zero to satisfy the non-negativity and integer constraints.

\paragraph{IV. Probabilistic top-down sampling}
We develop a sampling algorithm that takes samples from an upper level of the hierarchy and disaggregates them to generate coherent samples for the bottom series.
Instead of using fixed historical proportions as the traditional top-down  \citep{gross1990disaggregation}, our algorithm models historical proportions probabilistically as joint distributions.
We estimate these distributions in-sample, using Poisson or negative binomial distributed marginals and modeling cross-series dependence via copulas \citep{nelsen2006introduction}.
A detailed description of the sampling algorithm is provided in Sect.~\ref{sec:probTD}.

\paragraph{V. Aggregate joint bottom samples}
The samples for the bottom series obtained at the previous step are joint, meaning that they retain the dependence between forecasts of different series.
We thus obtain coherent forecasts of all aggregated series by
pre-multiplying by the aggregation matrix $\Abf$.

\subsection{Probabilistic top-down sampling}
\label{sec:probTD}

To simplify the exposition, we first describe the probabilistic top-down sampling algorithm on a minimal hierarchy, and then we show the extension to multiple bottom and upper series.

\paragraph{Minimal hierarchy}

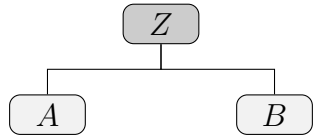
\begin{figure}[h!]
    \begin{center}
        \begin{tikzpicture}[every node/.style={rectangle,draw,align=center,rounded corners=4,font=\large},level distance=1.2cm,
  level 1/.style={sibling distance=3.cm, edge from parent fork down}
]
  \node[fill=black!20,minimum width=1cm] {$Z$}
    child {node[fill=black!5,minimum width=1cm] {$A$}}
    child {node[fill=black!5,minimum width=1cm] {$B$}};
\end{tikzpicture}
    \end{center}
    \caption{A minimal hierarchy with 1 upper and 2 bottom series.}
    \label{fig:simple_hier_TD}
\end{figure}

Let us first consider the hierarchy in Fig.~\ref{fig:simple_hier_TD} with one upper series $Z$ and two bottom series $A$ and $B$.
Given non-negative integer samples $z^{(1)}, \dots, z^{(N)}$ for time series $Z$ (obtained at step III),
our goal is to draw joint coherent samples 
\[
\big(a^{(1)},b^{(1)}\big), \dots, \big(a^{(N)},b^{(N)}\big)
\]
such that $a^{(i)}+b^{(i)} = z^{(i)}$, for all $i=1,\dots,N$.
For each $z^{(i)}$, we consider the finite set $\mathcal{S}\bigl(z^{(i)}\bigr)$ of all feasible pairs:
\begin{align}
\mathcal{S}\bigl(z^{(i)}\bigr)
&=\big\{(a,b):\ a,b\in\nn,\ a+b=z^{(i)}\big\} \nonumber \\
&=\big\{(0,z^{(i)}),\, (1,z^{(i)}\!-\!1),\, \dots,\, (z^{(i)},0)\big\}. \nonumber
\end{align}
We draw $(a^{(i)},b^{(i)})$ from $\mathcal{S}\bigl(z^{(i)}\bigr)$, using the probability distribution obtained by conditioning the joint distribution $\pihat_{A,B}$ of $A$ and $B$, estimated in-sample (see Sect.~\ref{sec:estimation_details} for details), on the constraint $A+B=z^{(i)}$. 
We compute the probability of any feasible pair $\bigl(j,\, z^{(i)}-j\bigr) \in \mathcal{S}\bigl(z^{(i)}\bigr)$, with $j=0,\dots,z^{(i)}$, as 
\[
p_j = \frac{\pihat_{A,B}\bigl(j,\, z^{(i)}-j\bigr)}{\sum_{k=0}^{z^{(i)}}\,\pihat_{A,B}\bigl(k,\, z^{(i)}-k\bigr)}.
\]

\begin{remark}
Given the upper-level samples $z^{(1)},\dots,z^{(N)}$, coherence is enforced by randomly splitting each $z^{(i)}$ across $A$ and $B$,
where the probability of each split is assigned according to the historical joint distribution $\pihat_{A,B}$ of bottom counts.
In this sense, $\pihat_{A,B}$ serves as a probabilistic counterpart of historical proportions in classical top-down methods. 
\end{remark}

\paragraph{Multiple bottom series}
Consider an upper series $Z$ with $m>2$ bottom descendants $X_1,\dots,X_m$.
We reduce the sampling problem to a sequence of bivariate splits, organized as a binary tree whose leaves are the bottom series $X_1,\dots,X_m$ and the root is $Z$. 
The tree is built by recursively partitioning the descendants: at each internal node the corresponding group of series is split into two disjoint subgroups $L$ and $R$ of similar size,
with aggregated series $X_L=\sum_{k\in L} X_k$ and $X_R=\sum_{k\in R} X_k$.
We then process the tree top-down, starting from the samples $z^{(i)}$ of $Z$. 
At each internal node we apply the bivariate top-down step of the minimal hierarchy: we estimate the joint distribution $\pihat_{X_L,X_R}$ and draw $\big(x_L^{(i)},\,x_R^{(i)}\big)$ from the conditional law; then, $x_L^{(i)}$ and $x_R^{(i)}$ become the totals to be disaggregated within (respectively) $L$ and $R$ at the next level. 
The recursion terminates at the bottom level; by
construction, for each $i$ we obtain a coherent vector
$\big(x_1^{(i)},\dots,x_m^{(i)}\big)$, whose sum equals $z^{(i)}$.

\begin{figure}[h!]
\centering

\begin{tikzpicture}[
  every node/.style={rectangle, draw, align=center, rounded corners=4, font=\small, minimum height=.6cm},
  level distance=1.8cm,
  level 1/.style={sibling distance=3cm, edge from parent fork down},
  level 2/.style={sibling distance=2cm, edge from parent fork down}
]
  \node[fill=yellow!30] {\textbf{Hobbies 1}}
    child {node {\textbf{Item 1}}}
    child[edge from parent/.style={draw=none}] {node[draw=none, minimum height=1.cm] {\large \textbf{\dots}}}
    child {node {\textbf{Item 416}}};
\end{tikzpicture}

\vspace{3 mm}
\begin{tikzpicture}[>=latex]
  \draw[double distance=1pt, -{Latex[length=3mm, width=2mm]}, thick] (0,0) -- (0,-1);
\end{tikzpicture}
\vspace{5 mm}

\hspace*{-3.3mm}
\begin{tikzpicture}[
  every node/.style={rectangle,draw,align=center,rounded corners=4,font=\small},
  level distance=1.8cm,
  level 1/.style={sibling distance=2.6cm,
                  level distance=1.5cm,
                  minimum height=.6cm,
                  edge from parent fork down},
  level 2/.style={sibling distance=2.cm,
                  level distance=1.cm,
                  edge from parent fork down},
  level 3/.style={sibling distance=2.cm,
                  level distance=1.6cm,
                  edge from parent fork down}
]
  \node[fill=yellow!30, minimum height=.6cm] {\textbf{Hobbies 1}}
    child {node {\textbf{Item 1} $+\dots+$ \textbf{Item 208}}
      child[edge from parent/.style={draw=none}] {node[draw=none, minimum height=1.cm] {}
        child[edge from parent/.style = {draw}] {node {\textbf{Item 1}}}
        child[edge from parent/.style = {draw}] {node {\textbf{Item 2}}}
      }
      child[edge from parent/.style={draw=none}] {node[draw=none, minimum height=1.cm] {\large \textbf{\dots}}}
      child[edge from parent/.style={draw=none}] {node[draw=none, minimum height=1.cm] {}}
    }
    child[edge from parent/.style={draw=none}] {node[draw=none] {}
      child[edge from parent/.style={draw=none}] {node[draw=none] {}
        child[edge from parent/.style={draw=none}] {node[draw=none] {\large \textbf{\dots}}}
        }
    }
    child {node {\textbf{Item 209} $+\dots+$ \textbf{Item 416}}
      child[edge from parent/.style={draw=none}] {node[draw=none, minimum height=1.cm] {}}
      child[edge from parent/.style={draw=none}] {node[draw=none, minimum height=1.cm] {\large \textbf{\dots}}}
      child[edge from parent/.style={draw=none}] {node[draw=none, minimum height=1.cm] {}
        child[edge from parent/.style = {draw}] {node {\textbf{Item 415}}}
        child[edge from parent/.style = {draw}] {node {\textbf{Item 416}}}
      }
    };
\end{tikzpicture}

\caption{
Recursive binary splitting of the 416 bottom series under \texttt{Hobbies~1}.
}
\label{fig:binary_tree}
\end{figure}
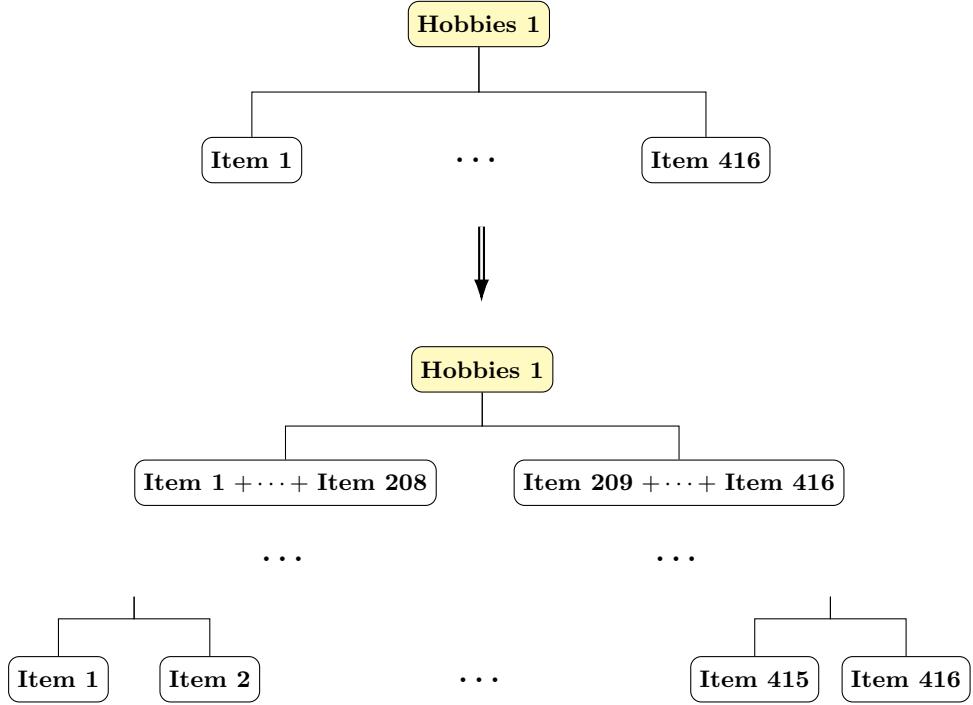

As an example, in the M5 hierarchy the series \texttt{Hobbies~1} is given by the sum of the bottom series $X_1,\dots,X_{416}$ corresponding to 416 different items (Fig.~\ref{fig:binary_tree}).
We first form the aggregated series $X_{1:208}=\sum_{j=1}^{208}X_j$ and $X_{209:416}=\sum_{j=209}^{416}X_j$,
estimate the joint distribution $\pihat_{1:208,\, 209:416}$,
and sample $\big(x_{1:208}^{(i)},\,x_{209:416}^{(i)}\big)$ conditionally on the sample $z^{(i)}$ from the series \texttt{Hobbies~1}. 
Next, we split $X_{1:208}$ into $X_{1:104}$ and $X_{105:208}$, estimate $\pihat_{1:104,\, 105:208}$, and sample $\big(x_{1:104}^{(i)},x_{105:208}^{(i)}\big)$ conditional on $x_{1:208}^{(i)}$; we do the same for $X_{209:416}$.
We continue recursively until reaching single items, yielding joint samples for all $416$ bottom series. 

\paragraph{Multiple lowest-upper series}
Let $Z_1,\dots,Z_l$ denote the time series at the lowest level of the upper subhierarchy selected in step~I,
and let $\big(z_1^{(i)},\dots,z_l^{(i)}\big)$ be the joint reconciled samples obtained in step~III.
We apply the probabilistic top-down sampling algorithm independently for each $j=1,\dots,l$,
taking care not to reshuffle the sample index $i$, so that cross-node dependence inherited from the reconciliation step is retained.
This yields, for each $i$, coherent bottom-level samples under every $Z_j$, while preserving the joint structure across all lowest-upper series.

\subsection{Estimation of the bivariate joint distributions}
\label{sec:estimation_details}

A key component of the top-down sampling algorithm is the in-sample estimation of the joint distribution $\pihat_{A,B}$ of each pair of series $(A,B)$, across which we disaggregate. 
We fit $\pihat_{A,B}$ to the historical observations of the pair, assuming the disaggregation proportions are approximately time-invariant. To allow for mild non-stationarity, we weight observations by an exponential decay with a half-life of 28 days (four weekly cycles) as a heuristic, which favors more recent observations.

We adopt a simple parametric approach,
estimating the  marginal distributions and the dependence structure separately.
For the marginals, we fit a Poisson or a negative binomial (NB) distribution, depending on whether the data are under- or over-dispersed.
Parameters are estimated by moment matching, using the recency weights described above.
Moment matching coincides with maximum likelihood (ML) in the Poisson case; in the NB case it provides a fast, closed-form approximation that avoids the iterative optimization required by ML \citep{cameron2013regression}.

\begin{figure}[h!]
    \centering

    \includegraphics[width=0.32\textwidth]{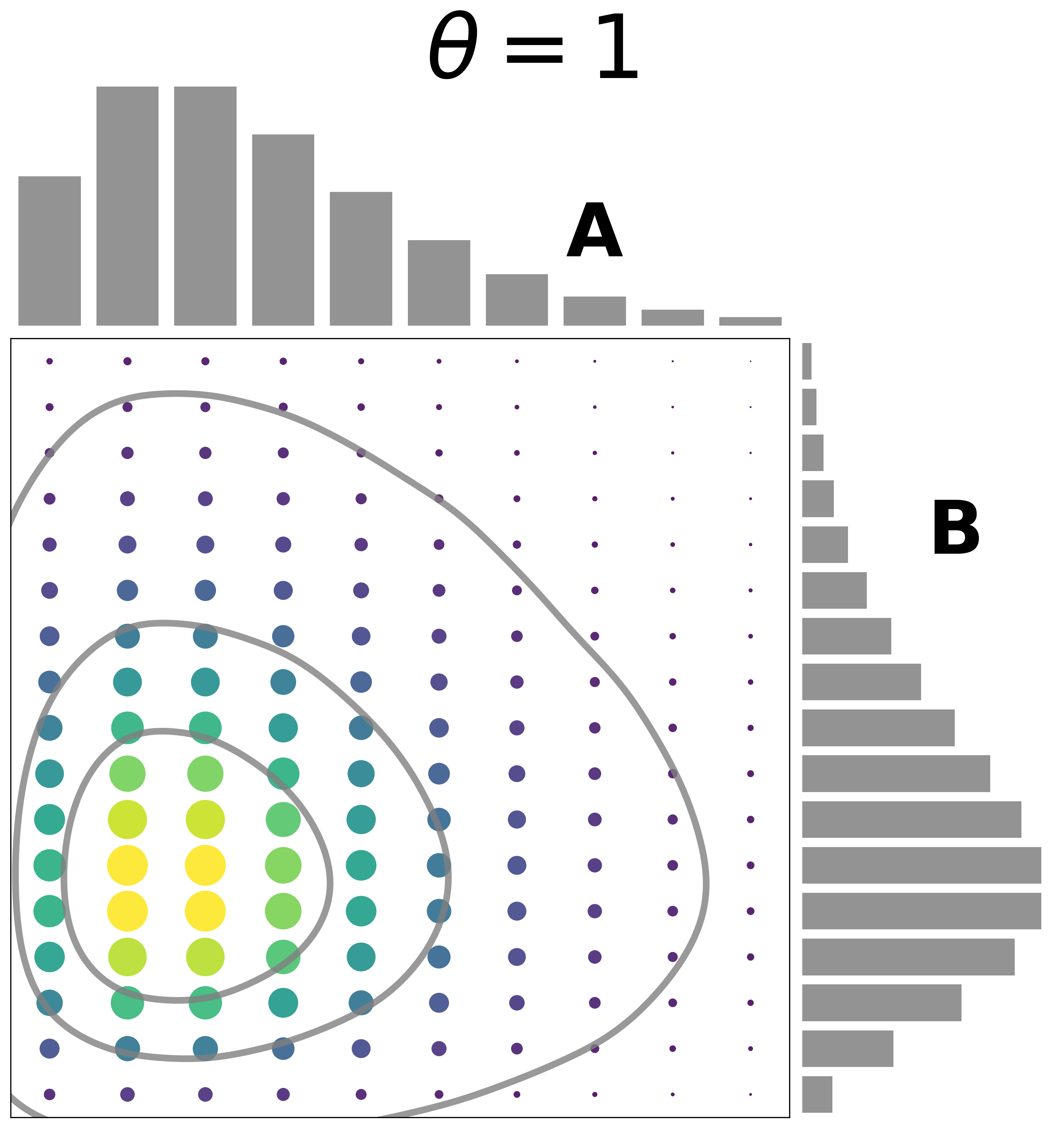} 
    \includegraphics[width=0.32\textwidth]{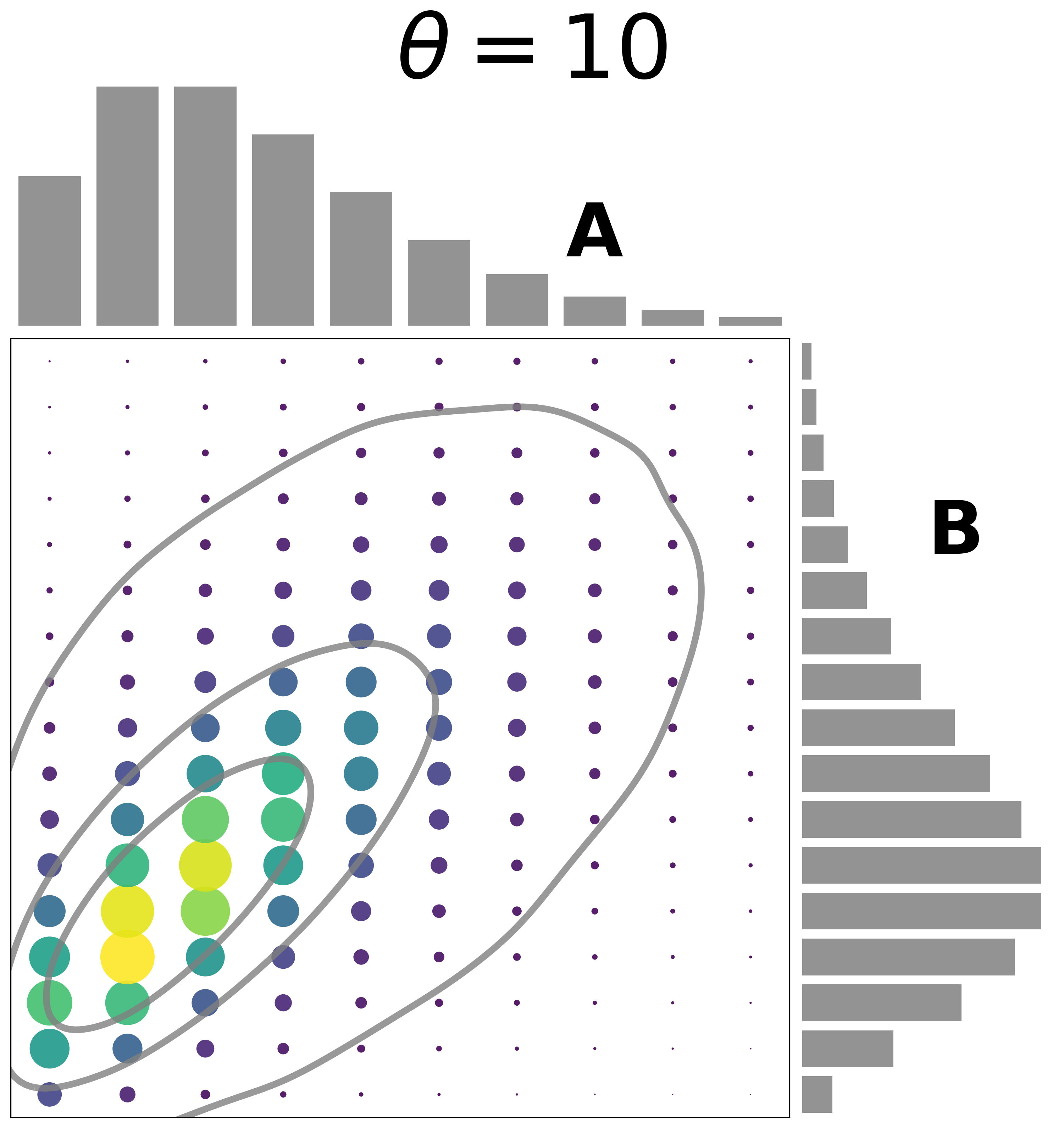}
    \includegraphics[width=0.32\textwidth]{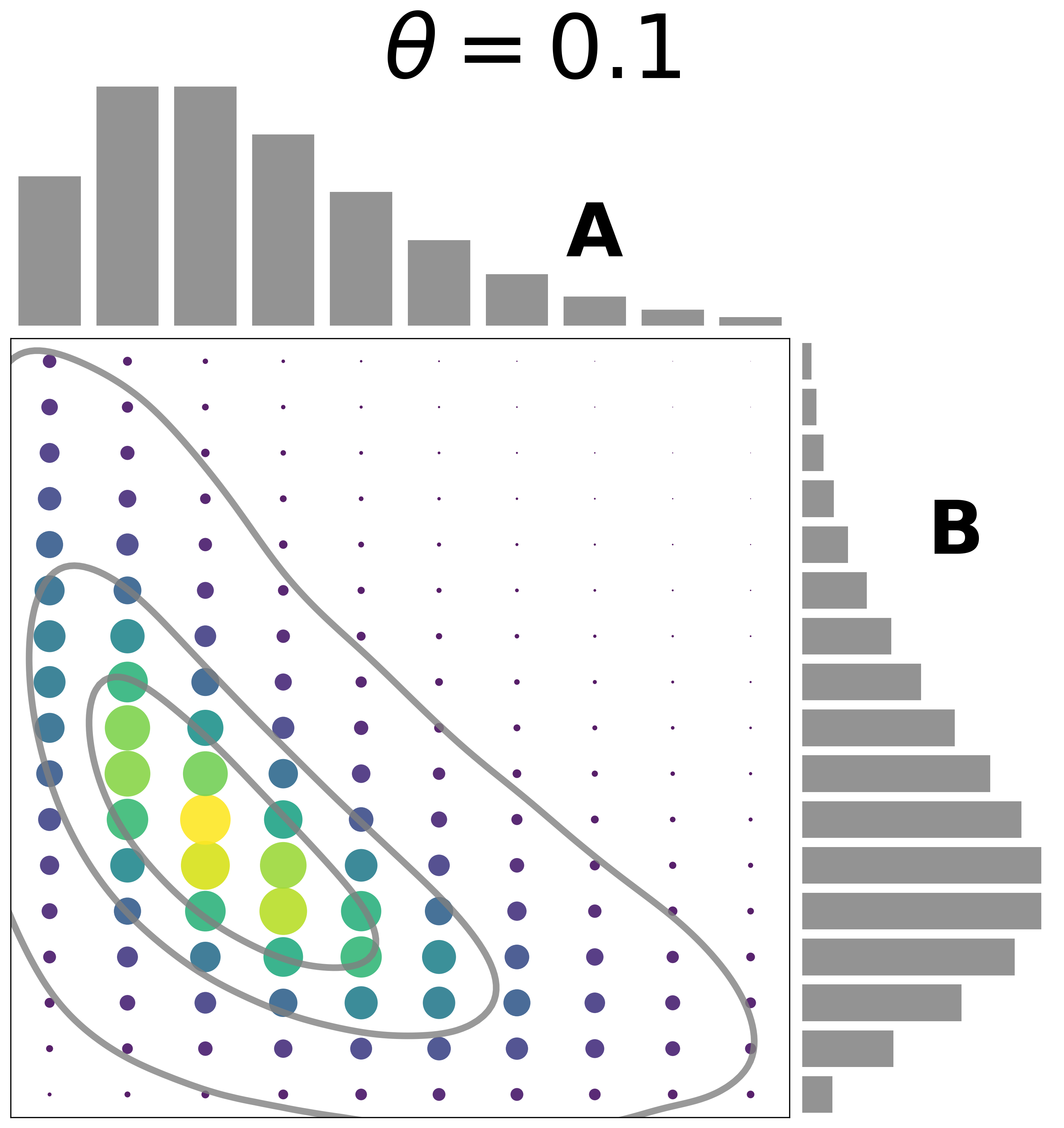}

    \caption{Joint distribution of the variables $A, B$. 
    The marginals are always distributed as $\NB(5, 0.5)$ and $\NB(10, 0.6)$ respectively. 
    The joint dependency is modeled via a Plackett copula with parameter $\theta$. 
    When $\theta = 1$, the variables are independent (left);
    a positive dependence corresponds to $\theta>1$ (center), a negative dependence to $\theta<1$ (right). 
    }
    \label{fig:ex_copula}
\end{figure}

We model the dependence between $A$ and $B$ with a copula \citep{sklar1959fonctions, nelsen2006introduction}, which couples the fitted marginals into a joint distribution.
Specifically, we use a bivariate Plackett copula, parametrized by a single parameter $\theta > 0$, which measures the strength of the dependence. 
Figure~\ref{fig:ex_copula} shows the joint distributions obtained by applying a Plackett copula with different values of $\theta$ to the same NB marginals.
We estimate $\theta$ by inverting the closed-form relationship between $\theta$ and Spearman's rank correlation $\rho$ \citep[Ch.~3]{nelsen2006introduction}: 
we compute $\rho$ from the data and solve numerically for the corresponding $\theta$.
As for the marginals, this is much faster than ML, as it replaces likelihood optimization with a single one-dimensional root-finding step.

We assess the speed of both estimators in a simulated benchmark ($T = 1941$, matching the M5 series length): 
moment matching and Spearman inversion are  roughly two orders of magnitude faster than ML, with near-identical estimates in both cases.
This is relevant at large scale, as \met performs about $n_b$ copula and $2n_b$ marginal estimations across the tree (one bivariate split for each internal node).
For M5 ($n_b \approx 30$K), ML-based estimation would take tens of minutes, against the few seconds required by the adopted fast estimators.

\subsection{Computational strategies}
\label{sec:computational_strategies}

Several implementation choices reduce computation time without affecting the resulting forecast distribution,
allowing \met to handle large hierarchies efficiently.

\paragraph{Batching over unique totals}
In the top-down sampling algorithm, we avoid looping over every total $z^{(i)}$ when sampling from the conditional distribution $\pihat_{A,B}$ given the constraint $A+B=z^{(i)}$ (see Sect.~\ref{sec:probTD}).
The probabilities of all feasible pairs in $\mathcal{S}\bigl(z^{(i)}\bigr)$ are computed once for each distinct integer value of $z^{(i)}$, and used to draw as many samples as there are occurrences of that value.
Because the $z^{(i)}$ are integer-valued and often small, they take few distinct values,
making this batching step particularly efficient. 
The sampled values are then written back to their original positions to preserve sample alignment and thus cross-series dependence.

\paragraph{Multi-step-ahead forecasts}
We produce forecasts for all horizons $h = 1, \dots, H$ in a single top-down pass rather than one horizon at a time.
The distributions $\pihat_{A,B}$ used in the top-down algorithm do not depend on $h$;
we thus estimate their parameters once and reuse them across all horizons.
The reconciled upper samples, by contrast, differ across horizons; we concatenate them into one set of $N \cdot H$ samples, where $N$ is the number of samples per horizon.
The top-down algorithm is then run a single time on the combined set, with the unique-totals batching (described above) applied to the pooled samples.

\paragraph{Parallelization}
Forecasting the selected upper series using univariate models is trivially parallelized across series.
The top-down sampling runs independently for each series in the lowest level of the upper subhierarchy and is thus parallelized across them, enabling efficient execution on multi-core hardware.

\paragraph{Implementation}
\met is implemented in an R package, which will be publicly released upon acceptance.
The code to reproduce the experiments in the paper is available at \url{https://github.com/anonymoussubmission593-hub/e2eTD}.
Operations involving the aggregation matrix rely on sparse linear algebra via the \texttt{Matrix} package \citep{Matrix_package}.
The most computationally intensive steps of the top-down algorithm, such as computing the conditional split probabilities, sampling from the resulting categorical distributions, and extracting quantiles from the samples, are written in C++ via the \texttt{Rcpp} package \citep{eddelbuettel2011rcpp}.

\subsection{Positioning of \met}

We clarify here how e2eTD differs from the most closely related methods that adopt a top-down approach.
Both \citet{das23dirichlet} and \citet{long2025} produce the forecasts from a single aggregation level (respectively, the top and an intermediate level) and are thus exposed to the information loss and estimation risk of single-level approaches \citep{athanasopoulos2024forecast};
\met instead computes forecasts for different aggregation levels and combines them through Gaussian reconciliation before disaggregating.
The two methods differ from \met in further aspects.
\citet{das23dirichlet} focus only on strictly hierarchical time series, and the continuous Dirichlet proportions are a less natural fit for the low-count bottom series typical of retail.
Moreover, they learn the proportions by training a deep neural network, whereas \met estimates them from in-sample joint distributions.
\citet{long2025}, on the other hand, disaggregate the point forecasts to set parametric forecast distributions independently for each bottom series, retaining no joint structure; 
\met instead disaggregates samples via a probabilistic top-down algorithm, preserving the joint structure and yielding coherent forecasts at every level.
Closest to \met is the top-down conditioning approach of \citet{zambon2024probabilistic}, which computes forecasts for all bottom series and combines them via convolution to build the bottom-up distributions for all upper nodes of the binary tree.
The independence assumption implied by the convolution constitutes a structural limitation: 
as the authors acknowledge, ignoring cross-series correlations yields bottom-up distributions that are too narrow.
In large hierarchies, this causes upper samples to fall outside the support, so that reconciliation breaks down; the implementation in the R package \texttt{bayesRecon} fails on both datasets considered in this work.
e2eTD instead estimates the joint distribution directly from the in-sample observations, avoiding the independence assumption and capturing cross-series dependence through a copula. 
Moreover, it does not require forecasting any bottom series, which makes it more scalable.

\section{Empirical evaluation}
\label{sec:emp_evaluation}
 
\paragraph{Datasets}
We consider two real-world retail datasets: \textit{M5} and \textit{Favorita}.
To our knowledge, these are the largest publicly available time series datasets in retail.

The \textit{M5} dataset \citep{makridakis2022m5} consists of daily unit sales for 3,049 products across 10 Walmart stores in the United States, yielding 30,490 bottom series.
Bottom series are organized along two crossed hierarchies: a geographic dimension (State -- Store) and a product dimension (Category -- Department -- Item); all cross-combinations form 12 aggregation levels (Table~\ref{tab:combined_hierarchical_structure}).
Training data span from January 29, 2011 to June 19, 2016, for a total of 1,941 daily observations; following the competition setup, the last 28 days are held out for evaluation.
The dataset includes two sets of covariates: a binary holiday/event indicator and a  Supplemental Nutrition Assistance Program (SNAP) indicator, which encodes participation in a US federal food assistance program. 

The \textit{Favorita} dataset \citep{corporacion_favorita_2018} provides daily unit
sales for a nationwide supermarket chain in Ecuador, organized into a geographic
hierarchy (State -- City -- Store) and a product hierarchy (Family -- Class -- Item).
We restrict the series to the period January--August 2017, as recommended in the
literature \citep{olivares2024probabilistic}, retaining 227 daily observations for training and using the last 28 days for evaluation.
Since the raw data contain continuous-valued observations (items sold by weight or volume), we apply a systematic integer-valued filtering procedure.
We remove any product class in which more than 50\% of items have any non-integer sale; for the remaining classes, non-integer items are discarded individually.
In addition, negative sales (product returns) are clamped to zero, and store--item pairs with no positive sales in the training window are excluded.
After preprocessing, the dataset retains 161,480 of the original 174,685 bottom series, hierarchically organized into 16 aggregation levels (Table~\ref{tab:combined_hierarchical_structure}).
Alongside the sales records, we use two sets of covariates provided by the dataset: a promotion indicator (whether each item was on promotion on a given day) and a holiday indicator, which records national, regional, and local holidays; we additionally construct a day-before-holiday indicator.

\begin{figure}[h!]
    \centering
    \includegraphics[width=0.8\textwidth]{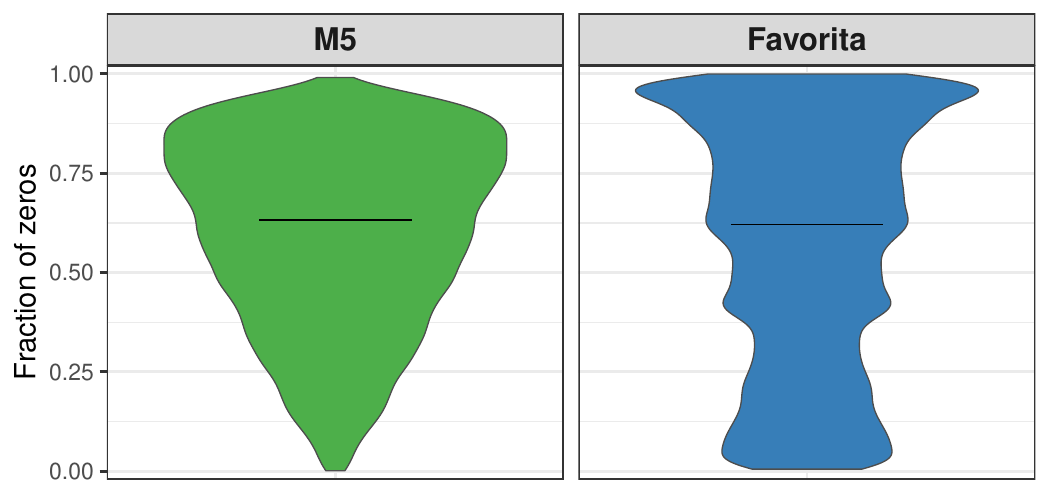}\\[-1ex]
    \includegraphics[width=0.8\textwidth]{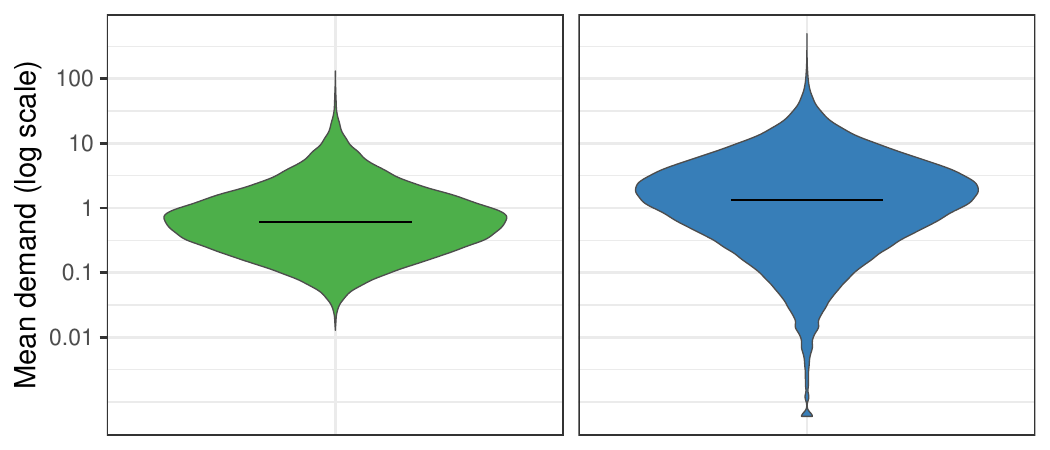}
    \caption{Violin plots representing the distributions of the fraction of zero observations (top) and mean demand (bottom) across bottom series, 
    computed on the training set of each dataset (left: \textit{M5}, right: \textit{Favorita}). 
    The scale of the mean demand is logarithmic for better visualization.
    The horizontal line corresponds to the median of the distribution.
    }
\label{fig:bottom_characteristics}
\end{figure}

Fig.~\ref{fig:bottom_characteristics} reports the distributions of the fraction of zeros and the mean demand per bottom series, computed on the training portion of each dataset.
The bottom series of \textit{M5} are more intermittent than those of \textit{Favorita}, 
which also exhibit higher mean demand.

\begin{table}[h!]
\centering
\caption{Hierarchical aggregation levels for the \textit{M5} and \textit{Favorita} datasets.
}
\label{tab:combined_hierarchical_structure}
\begin{tabular}{lllll}
\toprule
& \multicolumn{2}{c}{\textbf{M5}} & \multicolumn{2}{c}{\textbf{Favorita}} \\
\cmidrule(lr){2-3} \cmidrule(lr){4-5}
\textbf{Level} & \textbf{Aggregation} & \textbf{N. series} & \textbf{Aggregation} & \textbf{N. series} \\
\midrule
L1 & Total & 1 & Total & 1 \\
L2 & State & 3 & State & 16 \\
L3 & Store & 10 & City & 22 \\
L4 & Category & 3 & Store & 54 \\
L5 & Department & 7 & Family & 32 \\
L6 & State $\times$ Cat. & 9 & Class & 306 \\
L7 & State $\times$ Dept. & 21 & Item & 3,677 \\
L8 & Store $\times$ Cat. & 30 & State $\times$ Family & 501 \\
L9 & Store $\times$ Dept. & 70 & City $\times$ Family & 683 \\
L10 & Item & 3,049 & Store $\times$ Family & 1,670 \\
L11 & State $\times$ Item & 9,147 & State $\times$ Class & 4,486 \\
L12 & Store $\times$ Item & 30,490 & City $\times$ Class & 6,064 \\
L13 & & & Store $\times$ Class & 14,952 \\
L14 & & & State $\times$ Item & 49,847 \\
L15 & & & City $\times$ Item & 66,107 \\
L16 & & & Store $\times$ Item & 161,480 \\
\bottomrule
\end{tabular}
\end{table}



\paragraph{Forecasting model for the upper series}

The e2eTD method requires computing forecasts only on a selected subset of smooth aggregate series, as explained in Sect.~\ref{sec:e2etd}.
For the \textit{M5} dataset, we compute forecasts on series aggregated to at least the Department level, corresponding to levels L1--L9, for a total of 154 series (0.36\% of the total).
For the \textit{Favorita} dataset, we compute forecasts on series aggregated to at least the Family and City levels, corresponding to levels L1, L2, L3, L5, L8, and L9.
Some series are duplicated across levels; after excluding duplicates, we fit models on the remaining 848 unique series (0.28\% of the total).




%

We forecast the selected upper series using exponential smoothing (ETS) with automatic model selection \citep[Ch.~8]{hyndman2021fpp3}.
ETS is fit independently for each upper series, making the procedure embarrassingly parallel across cores.
ETS is generally considered a strong benchmark for aggregate time series in retail, which typically show a clear trend and seasonality.
It is commonly used by practitioners also for its simplicity, interpretability, and light computational cost \citep{wellens2024simplifying, martins2022retail}.
We use the implementation provided by the R package \texttt{smooth} \citep{smooth_pkg}, which supports exogenous regressors. This enables us to incorporate additional demand drivers beyond the sales signal itself.

The role of promotions in improving retail forecasts is well documented \citep{kourentzes2016forecasting}; we therefore include the promotion indicator as a regressor for \textit{Favorita}.
Walmart stores instead operate under an everyday-low-price strategy \citep{long2025}, so no promotional covariate is available. 
Nevertheless, we include the holiday/event and SNAP indicators; the latter is particularly relevant as benefit disbursement dates typically yield demand spikes.
Besides the provided promotion and holiday indicators, for \textit{Favorita}
we further include a payday regressor, reflecting the bi-monthly wage payment dates in Ecuador. 
%
Moreover, bottom-level series often exhibit leading zeros corresponding to products not yet on sale, rather than genuine zero demand \citep{makridakis2022m5}; the resulting aggregate series may thus display an artificial growth phase in their early observations.
To account for this, we include for each upper series an additional regressor counting the number of contributing bottom series that have already recorded at least one positive sale.
Finally, the ETS model naturally accounts for weekly seasonality; for \textit{M5}, whose training window spans nearly five years, we additionally include two pairs of Fourier terms to capture the annual seasonality \citep[Ch.~12]{svetunkov2023forecasting}.

\paragraph{Competing methods}

We compare e2eTD against five competing methods that produce coherent probabilistic forecasts.
Among them, we restrict to methods that scale to hierarchies with hundreds of thousands of series and can be run on a laptop; 
all the considered methods complete in under 6 minutes on \textit{M5} and 40 minutes on \textit{Favorita} (see Sect.~\ref{sec:comp_times}).
This excludes end-to-end neural approaches such as those by \citet{rangapuram2021end} and \citet{das23dirichlet}, as well as reconciliation methods that require a dense covariance structure \citep{Wickramasuriya2019}.
We could not apply the reconciliation methods for mixed-type hierarchies implemented in the \texttt{bayesRecon} R package \citep{bayesRecon}, as they produced runtime errors related to numerical instabilities.
The five methods considered are described below.

\begin{itemize}
    \item \textbf{empD}: empirical distribution. 
    For each series, we estimate the forecast distribution non-parametrically as the empirical distribution of the in-sample observations.
    Note that, since the training data are coherent by construction, the resulting joint empirical distribution is also coherent.
    This method corresponds to the Kernel benchmark of the M5 Uncertainty competition \citep{makridakis2022m5uncertainty}.
    \item \textbf{S-empD}: seasonal empirical distribution.
    A seasonal variant of empD in which, for each horizon $h$, the empirical distribution is computed using only the training observations falling on the same position within the seasonal cycle, i.e., on the same weekday.
    This reflects the strong weekly seasonality typical of retail data, with systematic differences in demand across weekdays \citep{kolassa2016evaluating}.
    \item \textbf{WLS}: seasonal naive with weighted least squares reconciliation.
    Since we need to compute the base forecasts for hundreds of thousands of series in the hierarchy, we adopt seasonal naive, instead of methods such as ETS that require optimization of the parameters.
    We scale the forecast variance proportionally to the number of elapsed seasonal cycles, following the M5 Uncertainty benchmark specification \citep{makridakis2022m5uncertainty}.
    Full MinT reconciliation \citep{Wickramasuriya2019} with a non-diagonal covariance matrix is computationally infeasible at the scale of \textit{M5} and \textit{Favorita}. 
    We therefore apply weighted least squares (WLS) reconciliation with a diagonal covariance, which retains the variance-scaling structure while remaining tractable.
    To make it scalable, we implement WLS using sparse linear algebra via the R package \texttt{Matrix} \citep{Matrix_package}.
    \item \textbf{Long}: method introduced by \citet{long2025} with bottom-up reconciliation.
    We fit the same ETS models used by e2eTD at the lowest level of the selected upper subhierarchy.
    We then obtain the means of the forecast distributions of the bottom series via a classical top-down disaggregation using historical proportions.
    We assume the bottom-level predictive distributions to be negative binomial, with 
    variance estimated from the in-sample observations; for series where the in-sample variance is smaller than the forecast mean, we use a Poisson distribution.
    Since the original method produces only bottom-level forecasts, we extend it to upper levels by bottom-up aggregation of independent samples drawn from the bottom-level distributions.
    \item \textbf{HINT}: hierarchical coherent network with mixture output \citep{olivares2023hint}.
    HINT combines a neural forecasting backbone (we use NHITS, \citealp{challu2023nhits}) with a mixture output layer and bottom-up sample reconciliation, which makes the resulting probabilistic forecasts hierarchically coherent by construction.
    The model is trained on the full hierarchy, using the same exogenous regressors as e2eTD; we use $K=10$ mixture components, as suggested by the authors.
    We replace the Gaussian mixture of the original paper with a Poisson mixture, which is better suited to count-valued series and corresponds to the output layer originally proposed for hierarchical count forecasting by \citet{olivares2024probabilistic}.
    In preliminary experiments, we also tested the Gaussian mixture, but it yielded similar results on \textit{Favorita} and considerably worse on \textit{M5}.
    Our implementation builds on the \texttt{neuralforecast} Python library \citep{olivares2022library_neuralforecast}, with a custom sparse bottom-up step required to scale to the size of \textit{Favorita}.  
\end{itemize}

\paragraph{Evaluation}

We evaluate probabilistic forecasts using the Weighted Scaled Pinball Loss (WSPL), following the evaluation scheme of the M5 Uncertainty competition \citep{makridakis2022m5}.
The pinball loss at quantile level $\alpha \in (0,1)$ for a forecast
$\hat{q}_\alpha$ and realization $y$ is
\begin{equation}
    \mathrm{PL}(\hat{q}_\alpha, y) =
    \begin{cases}
        \alpha\,(y - \hat{q}_\alpha) & \text{if } y \geq \hat{q}_\alpha, \\
        (1-\alpha)\,(\hat{q}_\alpha - y) & \text{otherwise.}
    \end{cases}
\end{equation}
Forecasts are evaluated at the nine quantile levels used in the M5 competition: $\alpha \in \{0.005, 0.025, 0.165, 0.25, 0.5, 0.75, 0.835, 0.975, 0.995\}$.
For each series, the raw pinball losses are averaged across quantile levels and forecast horizons.
The loss of each series is scaled by dividing by the in-sample mean absolute error of the one-step-ahead naive forecast.
The scaled losses are then aggregated across series to produce a WSPL score for each aggregation level using weights.
For \textit{M5} we adopt the competition weights, which are proportional to each series' total dollar sales revenue, so that high-revenue products contribute more to the overall score; for \textit{Favorita}, where no such weighting scheme is available, we use uniform weights.

\begin{table}[h!]
\centering
\caption{WSPL by aggregation level -- M5}
\label{tab:all_levels_m5}
\begin{tabular}{lcccccc}
\toprule
Level & e2eTD & empD & S-empD & WLS & Long & HINT \\
\midrule
L1 & \textbf{0.074} & 0.502 & 0.509 & 0.241 & 0.122 & 0.268 \\
L2 & \textbf{0.100} & 0.465 & 0.449 & 0.277 & 0.175 & 0.274 \\
L3 & \textbf{0.119} & 0.484 & 0.473 & 0.248 & 0.174 & 0.278 \\
L4 & \textbf{0.095} & 0.486 & 0.475 & 0.257 & 0.140 & 0.271 \\
L5 & \textbf{0.120} & 0.502 & 0.477 & 0.283 & 0.178 & 0.293 \\
L6 & \textbf{0.121} & 0.451 & 0.425 & 0.290 & 0.178 & 0.275 \\
L7 & \textbf{0.146} & 0.458 & 0.428 & 0.300 & 0.202 & 0.286 \\
L8 & \textbf{0.141} & 0.465 & 0.444 & 0.254 & 0.182 & 0.278 \\
L9 & \textbf{0.171} & 0.465 & 0.439 & 0.271 & 0.208 & 0.283 \\
L10 & 0.323 & 0.438 & 0.429 & 0.395 & 0.386 & \textbf{0.312} \\
L11 & 0.288 & 0.376 & 0.370 & 0.378 & 0.330 & \textbf{0.288} \\
L12 & \textbf{0.272} & 0.312 & 0.310 & 0.410 & 0.306 & 0.289 \\
\midrule
Mean & \textbf{0.164} & 0.450 & 0.436 & 0.300 & 0.215 & 0.283 \\
\bottomrule
\end{tabular}

\end{table}

\subsection{Forecast accuracy across the hierarchy}

\begin{table}[h!]
\centering
\caption{WSPL by aggregation level -- Favorita}
\label{tab:all_levels_favorita}
\begin{tabular}{lcccccc}
\toprule
Level & e2eTD & empD & S-empD & WLS & Long & HINT \\
\midrule
L1 & \textbf{0.112} & 0.159 & 0.114 & 0.192 & 0.207 & 0.207 \\
L2 & \textbf{0.143} & 0.196 & 0.166 & 0.227 & 0.235 & 0.231 \\
L3 & \textbf{0.146} & 0.198 & 0.169 & 0.229 & 0.242 & 0.231 \\
L4 & \textbf{0.167} & 0.199 & 0.173 & 0.222 & 0.269 & 0.233 \\
L5 & \textbf{0.124} & 0.168 & 0.129 & 0.196 & 0.206 & 0.216 \\
L6 & \textbf{0.167} & 0.197 & 0.172 & 0.214 & 0.207 & 0.244 \\
L7 & \textbf{0.225} & 0.249 & 0.241 & 0.277 & 0.242 & 0.292 \\
L8 & \textbf{0.174} & 0.210 & 0.185 & 0.228 & 0.237 & 0.245 \\
L9 & \textbf{0.177} & 0.212 & 0.189 & 0.229 & 0.240 & 0.245 \\
L10 & \textbf{0.196} & 0.218 & 0.200 & 0.233 & 0.257 & 0.254 \\
L11 & \textbf{0.202} & 0.235 & 0.221 & 0.252 & 0.237 & 0.276 \\
L12 & \textbf{0.203} & 0.236 & 0.223 & 0.252 & 0.238 & 0.278 \\
L13 & \textbf{0.218} & 0.238 & 0.231 & 0.277 & 0.243 & 0.290 \\
L14 & \textbf{0.222} & 0.242 & 0.242 & 0.300 & 0.238 & 0.326 \\
L15 & \textbf{0.225} & 0.244 & 0.244 & 0.311 & 0.240 & 0.363 \\
L16 & \textbf{0.231} & 0.238 & 0.244 & 0.375 & 0.242 & 0.504 \\
\midrule
Mean & \textbf{0.183} & 0.215 & 0.196 & 0.251 & 0.236 & 0.277 \\
\bottomrule
\end{tabular}

\end{table}

Table~\ref{tab:all_levels_m5} and Table~\ref{tab:all_levels_favorita} report the WSPL at each aggregation level and its mean across levels for \textit{M5} and \textit{Favorita}, respectively.
On both datasets, e2eTD achieves the best mean WSPL by a substantial margin, with a clearer improvement on \textit{M5} than on \textit{Favorita}.
The advantage is clear and consistent across most aggregation levels, indicating that the method produces well-calibrated forecasts throughout the entire hierarchy and not only at the levels where the forecasts are directly computed (L1--L9 for \textit{M5}; L1, L2, L3, L5, L8, L9 for \textit{Favorita}).
On \textit{M5}, the performance of e2eTD can also be compared against the results of the M5 Uncertainty competition \citep{makridakis2022m5uncertainty}: e2eTD would have ranked 11th out of 892 teams. 
On \textit{Favorita}, no such comparison is possible, as the corresponding competition evaluated only point forecasts.

The simple empirical baselines, empD and S-empD, are competitive at the bottom level on both datasets, confirming previous results in the literature \citep{spiliotis2021product, kolassa2016evaluating}.
S-empD consistently outperforms empD, confirming that weekly seasonality is an important driver of retail demand.
On \textit{M5}, however, both baselines perform poorly at upper aggregation levels: their static empirical distribution does not account for trends and temporal dynamics, which become more pronounced at higher levels of aggregation and over the long M5 training window.
On \textit{Favorita}, S-empD is notably competitive (second-best method after e2eTD); a likely reason is the shorter training period and the strong weekly seasonality that dominates aggregate demand.
WLS performs poorly on both datasets.
Multiple factors contribute to this: the seasonal naive base forecasts are weak; the Gaussian distributional assumption is poorly suited to the lower aggregation levels, where many series are intermittent or consist of low counts; and the diagonal covariance structure ignores cross-series correlations.
The Long method was originally designed to produce only bottom-level forecasts \citep{long2025}. 
To obtain coherent upper forecasts, we aggregate independent samples drawn from the bottom-level distributions; 
ignoring the cross-series dependence causes its performance to degrade substantially at upper aggregation levels.
This is most visible on \textit{M5}: the WSPL is about 30\% larger than that of e2eTD if we consider all aggregation levels, but only 12\% if we consider just the bottom level.
Finally, HINT shows an asymmetric behavior across the two datasets.
On \textit{M5}, HINT is the third-best method overall; it is particularly effective on low aggregation levels, ranking first on L10 and L11, but its performance degrades at the upper levels.
On \textit{Favorita}, however, HINT ranks last, with poor performance also at the lower aggregation levels.
A possible explanation is that the larger number of series in \textit{Favorita}, combined with the shorter training period, makes it harder to fit the underlying neural network.

\begin{figure}[h!]
    \centering
    \includegraphics[width=0.99\linewidth]{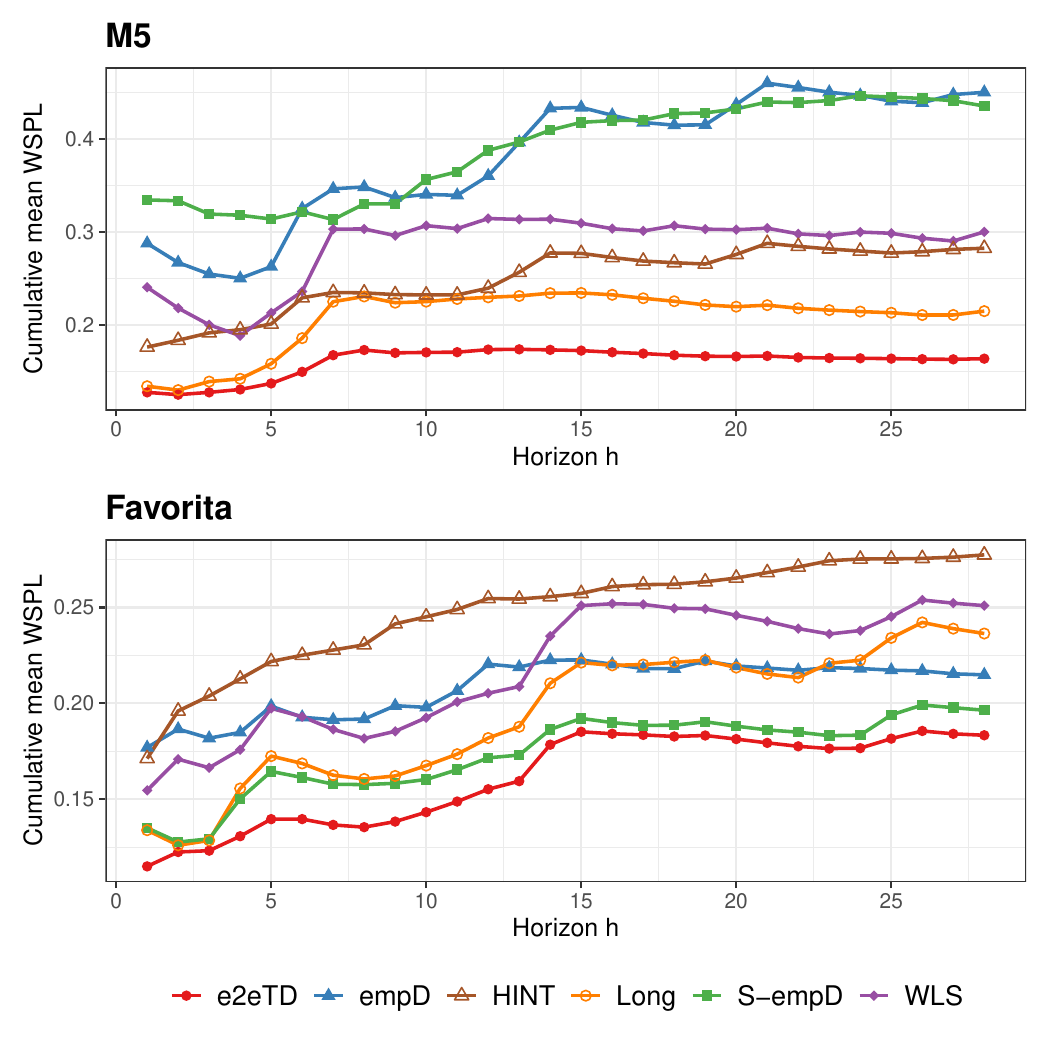}
    \caption{Cumulative mean WSPL as a function of the forecast horizon $h$, averaged over all aggregation levels, for \textit{M5} (top) and \textit{Favorita} (bottom).}
    \label{fig:wspl_vs_h}
\end{figure}

In Fig.~\ref{fig:wspl_vs_h}, we show the cumulative mean WSPL as a function of the forecast horizon $h$, for both datasets.
e2eTD achieves the lowest WSPL at every horizon on both datasets, with all curves rising over the first few days and stabilizing after one to two weeks.
Long performs comparably to e2eTD at short horizons, especially on \textit{M5}, but deteriorates more rapidly as $h$ grows.
A plausible explanation is that Long relies on a static, in-sample estimate of the bottom-level variance, whereas e2eTD inherits a horizon-dependent predictive variance from the ETS forecasts of the upper series.

\subsection{Bottom-level accuracy}


\begin{table}[h!]
\centering
\caption{Bottom-level WSPL by quantile level}
\label{tab:bott_q}
\begin{tabular}{llrrrrrr}
\toprule
Dataset & Quantile & e2eTD & empD & S-empD & WLS & Long & HINT \\
\midrule
\multirow{4}{*}{M5} & $0.750$ & \textbf{0.528} & 0.652 & 0.644 & 0.666 & 0.635 & 0.562 \\
 & $0.835$ & \textbf{0.435} & 0.550 & 0.542 & 0.546 & 0.536 & 0.472 \\
 & $0.975$ & \textbf{0.126} & 0.158 & 0.160 & 0.162 & 0.158 & 0.143 \\
 & $0.995$ & \textbf{0.038} & 0.045 & 0.050 & 0.058 & 0.046 & 0.042 \\
\midrule
\multirow{4}{*}{Favorita} & $0.750$ & \textbf{0.450} & 0.470 & 0.467 & 0.627 & 0.473 & 0.937 \\
 & $0.835$ & \textbf{0.415} & 0.436 & 0.439 & 0.559 & 0.436 & 0.938 \\
 & $0.975$ & \textbf{0.213} & 0.224 & 0.243 & 0.235 & 0.225 & 0.524 \\
 & $0.995$ & 0.128 & \textbf{0.125} & 0.161 & 0.135 & 0.126 & 0.164 \\
\bottomrule
\end{tabular}

\end{table}

Table~\ref{tab:bott_q} reports the bottom-level WSPL at four representative quantiles.
We focus on upper quantiles, which are the most relevant ones for SKU-level series: such series are typically low-count and often intermittent, so practitioners are mainly interested in the upper tail of the predictive distribution for inventory and safety-stock decisions \citep{boylan2006accuracy, spiliotis2021product}.
e2eTD is the best method at $\alpha=0.75, 0.835, 0.975$ on both datasets, but the gap to the other methods narrows at $\alpha=0.995$: e2eTD still wins on \textit{M5}, while empD is marginally better on \textit{Favorita}.
This aligns with \citet{spiliotis2021product}, who find that the advantage of the winning methods over empD shrinks at higher quantiles; more generally, empirical baselines are notoriously hard to beat for low-count, intermittent series \citep{kolassa2016evaluating}.
WLS is comparable to the empirical baselines on \textit{M5} but performs generally worse on \textit{Favorita}, possibly due to the larger hierarchy.
Long is competitive at the bottom level, performing slightly better than the empirical baselines across quantiles on both datasets.
Finally, HINT exhibits the same asymmetric behavior observed in the main results: on \textit{M5} it is the second-best method at every quantile, while it ranks last on \textit{Favorita} by a wide margin, especially at the central upper quantiles.

\subsection{Computational cost}
\label{sec:comp_times}

\begin{table}[h!]
\centering
\caption{Computational times (minutes). The statistics are computed over 7 independent runs.
All methods run on CPU, except for HINT, which uses a GPU for training.
}
\label{tab:comp_times}
\begin{tabular}{lcccc}
\toprule
\multirow{2}{*}{} & \multicolumn{2}{c}{M5} & \multicolumn{2}{c}{Favorita} \\
\cmidrule(lr){2-3} \cmidrule(lr){4-5}
 & median & mean $\pm$ sd & median & mean $\pm$ sd \\
\midrule
e2eTD & 4.45 & 4.32 $\pm$ 0.33 & 17.38 & 19.91 $\pm$ 5.06 \\
empD & 0.13 & 0.13 $\pm$ 0.01 & 0.45 & 0.53 $\pm$ 0.20 \\
S-empD & 1.68 & 1.69 $\pm$ 0.03 & 10.31 & 10.52 $\pm$ 0.43 \\
WLS & 0.47 & 0.48 $\pm$ 0.02 & 40.02 & 40.03 $\pm$ 0.32 \\
Long & 5.99 & 5.97 $\pm$ 0.14 & 30.24 & 30.70 $\pm$ 1.63 \\
HINT & 1.99 & 2.06 $\pm$ 0.12 & 3.83 & 3.82 $\pm$ 0.05 \\
\bottomrule
\end{tabular}

\end{table}

Table~\ref{tab:comp_times} reports the running times (median and mean $\pm$ standard deviation over 7 independent runs) for each method on the two datasets, measured on a standard laptop\footnote{Hardware: 12th Gen Intel Core i7 CPU, 64 GB RAM, NVIDIA T600 GPU (4 GB VRAM). All methods run on CPU only, except HINT, which uses the GPU for neural network training.
ETS fitting and the top-down sampling step of e2eTD are run in parallel on 8 cores.}.
e2eTD takes about 4 minutes on \textit{M5} and 17 minutes on \textit{Favorita}; a detailed breakdown of its computational cost is reported below.
The empirical baselines empD and S-empD are the fastest methods overall, although on the larger \textit{Favorita} even S-empD takes about 10 minutes.
WLS is very fast on \textit{M5}, but becomes the slowest method on \textit{Favorita}, as the reconciliation step scales poorly with the size of the hierarchy.
Long is slightly slower than e2eTD; a substantial part of the cost is due to drawing the bottom-level samples from the estimated parametric distributions, aggregating them to obtain the upper-level samples, and extracting the quantiles. Note that these operations are not performed by the original method \citep{long2025}, which only produces bottom-level forecasts.
Finally, HINT is very fast (it is the second-fastest method on \textit{Favorita}); however, the comparison is not entirely fair, as HINT is the only method to use a GPU.
For both datasets, the running time of each method is well below one hour, which makes the comparison feasible on a single laptop.

\begin{table}[h!]
\centering
\caption{Computational times of the steps of e2eTD (seconds). The statistics are computed over 7 independent runs.
}
\label{tab:comp_times_e2eTD}
\begin{tabular}{lcccc}
\toprule
\multirow{2}{*}{} & \multicolumn{2}{c}{M5} & \multicolumn{2}{c}{Favorita} \\
\cmidrule(lr){2-3} \cmidrule(lr){4-5}
 & median & mean $\pm$ sd & median & mean $\pm$ sd \\
\midrule
\makecell[l]{computation of \\ upper forecasts} & 192.8 & 185.3 $\pm$ 15.9 & 107.8 & 112.6 $\pm$ 12.5 \\ \addlinespace[.3em]
\makecell[l]{reconciliation of \\ upper forecasts} & 0.50 & 0.50 $\pm$ 0.03 & 14.8 & 15.5 $\pm$ 1.6 \\ \addlinespace[.3em]
\makecell[l]{probabilistic \\ top-down sampling} & 51.9 & 53.2 $\pm$ 4.3 & 632.0 & 722.0 $\pm$ 178.5 \\ \addlinespace[.3em]
\makecell[l]{bottom-up aggregation \\ + quantile extraction} & 18.3 & 18.0 $\pm$ 1.1 & 286.1 & 343.0 $\pm$ 113.0 \\
\bottomrule
\end{tabular}

\end{table}

Table~\ref{tab:comp_times_e2eTD} breaks down the cost of e2eTD into its four main steps, corresponding to steps II--V of Fig.~\ref{fig:e2etd}; the last row also includes the computation of the quantiles from the forecast samples, which is non-trivial in large dimension.
Step I does not have any computational cost, as we decide which time series to forecast without using an automatic criterion.
On \textit{M5}, most of the runtime ($\sim$70\%) is spent fitting the ETS models to produce the upper-level forecasts.
On \textit{Favorita}, the cost profile is different: the upper forecasts take about two minutes (less than on \textit{M5}, due to the shorter training window), while the top-down sampling is the most expensive step, requiring more than 10 minutes.
The reconciliation step is negligible on both datasets, confirming that the analytic reconciliation embedded in e2eTD does not introduce a computational bottleneck.

\subsection{Ablation study}
\label{sec:ablation}

\begin{table}[h!]
\centering
\small
\caption{Ablation study: relative change in WSPL of each variant with respect to the default e2eTD configuration.}
\label{tab:ablation_compact}
\begin{tabular}{llccccc}
\toprule
Dataset & Level & indep. copula & no reconc. & small subhier. & no xreg & arima \\
\midrule
\multirow{2}{*}{M5} & L12 & +0.4\% & +0.6\% & +0.3\% & +0.7\% & +0.1\% \\
 & Mean & -0.6\% & +8.8\% & +2.8\% & +18.3\% & +1.7\% \\
\midrule
\multirow{2}{*}{Favorita} & L16 & +2.4\% & +0.3\% & +0.0\% & +0.1\% & +0.0\% \\
 & Mean & +4.7\% & +7.3\% & +0.9\% & +0.7\% & +1.7\% \\
\bottomrule
\end{tabular}

\end{table}

To assess the contribution of the individual design choices in e2eTD, we compare the default configuration against five ablation variants, each obtained by replacing a single component while keeping all others fixed:
\begin{itemize}
    \item \textit{indep. copula}: uses an independent copula instead of Plackett in the probabilistic top-down sampling (step IV);
    \item \textit{no reconc.}: the upper forecasts are not reconciled (step III) before the top-down step;
    \item \textit{small subhier.}: a smaller number of upper series is selected in step I;
    specifically, we use levels L1, L2, L3, L4, L6, L8 for \textit{M5} (56 series)
    and levels L1, L2, L5, L8 for \textit{Favorita} (541 series);
    \item \textit{no xreg}: no exogenous regressors are used to compute the upper forecasts (step II);
    \item \textit{arima}: upper forecasts are computed with ARIMA \citep[Ch.~9]{hyndman2021fpp3} rather than ETS, using the implementation provided by the R package \texttt{forecast} \citep{hyndman2008automatic}.
\end{itemize}
Table~\ref{tab:ablation_compact} reports the relative change in bottom-level WSPL and in mean WSPL across levels for each variant with respect to the default configuration; absolute values and detailed per-level results are reported in \ref{appendix:additional_results}.
Modeling dependences via a Plackett copula brings a clear improvement on \textit{Favorita} but essentially no improvement on \textit{M5}.
A plausible explanation is that cross-series dependence is stronger on \textit{Favorita}, while on \textit{M5} much of this dependence is already explained by the exogenous regressors at upper levels.
%
The reconciliation step on the upper forecasts yields a consistent improvement across the two datasets, confirming the well-known beneficial effect of combining information across aggregation levels \citep{athanasopoulos2024forecast}.
Using a smaller upper subhierarchy yields a slightly larger WSPL.
The size of the subhierarchy offers a tunable trade-off between computational cost and accuracy: more upper series are expected to provide better performance at the price of additional forecasting effort. 
Including exogenous regressors in the model for the upper forecasts helps where informative covariates are available, with a clear benefit on \textit{M5} and a negligible effect on \textit{Favorita}.
%
Finally, the choice between ETS and ARIMA has a marginal impact; on both datasets, ETS yields a small improvement in mean WSPL. 
We adopt ETS as the default because it is generally faster to fit; the ablation confirms that ARIMA is a comparable alternative.

\section{Discussion and conclusions}
\label{sec: conclusions}

We introduced e2eTD, a scalable method for coherent probabilistic forecasting of large hierarchical and grouped time series with intermittent series at the bottom level. 
e2eTD directly forecasts only a small subset of smooth upper series, combines them via standard reconciliation, and propagates the reconciled samples to the bottom level through a novel probabilistic top-down sampling algorithm.
The resulting bottom-level samples can be summed to obtain coherent probabilistic forecasts across all aggregation levels.
Coherence ensures that decisions taken at different levels rest on a consistent view of future demand, and the probabilistic representation supports risk-aware decisions such as setting safety stocks.
Empirically, e2eTD achieves the best mean WSPL across aggregation levels among all competing methods on the M5 and Favorita datasets, and would have ranked 11th out of 892 teams in the M5 Uncertainty competition \citep{makridakis2022m5uncertainty}. 
Notably, it does not directly forecast the bottom series, which constitute the majority of the hierarchy but typically carry little signal.
Yet e2eTD delivers strong accuracy at the bottom level, which drives inventory and replenishment decisions in retail \citep{fildes2022retail}: its bottom-level (L12) WSPL would have placed it 7th among the top 50 teams of the M5 Uncertainty competition.

Computational cost is a major concern at retail scale,
where even modest reductions in forecast cost translate into substantial financial and environmental savings \citep{petropoulos2025wielding}. 
e2eTD is designed to scale to large hierarchies,
as it directly forecasts only a small subset of the hierarchy (about $0.3\%$ of the total in our experiments);
the size of this subset can be tuned to control the trade-off between accuracy and computational cost.
Both the computation of the upper-level forecasts and the top-down sampling parallelize trivially across cores.
In our experiments, e2eTD runs in under 5 minutes on \textit{M5} and under 20 minutes on \textit{Favorita} on a standard laptop, with no specialized hardware required.

Several directions remain open for future work.
A first direction concerns the choice of upper-level series to forecast: principled criteria based on available computational resources, accuracy–cost trade-offs, or recently proposed forecastability measures \citep{wang2025time} could replace the manual selection used here. 
A second direction concerns the modelling of disaggregation proportions.
Our implementation uses a simple weighting heuristic that gives more importance to recent observations, accounting for slow drift in the data; more principled treatments could account for trend, seasonality, or time-varying parameters. 
A more substantial extension would replace historical proportions with forecast proportions \citep{Athanasopoulos2009tourism} for those series where temporal dynamics carry strong signal, together with a formal criterion to decide which bottom series to forecast.


\bibliographystyle{elsarticle-harv}
\bibliography{biblio}

\newpage
\appendix

\section{Additional results}
\label{appendix:additional_results}

\begin{table}[h!]
\centering
\caption{Ablation study - M5}
\label{tab:ablation-M5}
\begin{tabular}{lcccccc}
\toprule
Level & default & indep. copula & no reconc. & small subhier. & no xreg & arima \\
\midrule
L1 & 0.074 & 0.074 & 0.092 & 0.077 & 0.128 & 0.080 \\
L2 & 0.100 & 0.100 & 0.130 & 0.101 & 0.145 & 0.102 \\
L3 & 0.119 & 0.119 & 0.137 & 0.120 & 0.159 & 0.123 \\
L4 & 0.095 & 0.095 & 0.110 & 0.096 & 0.137 & 0.098 \\
L5 & 0.120 & 0.120 & 0.138 & 0.143 & 0.160 & 0.123 \\
L6 & 0.121 & 0.121 & 0.141 & 0.122 & 0.159 & 0.125 \\
L7 & 0.146 & 0.146 & 0.163 & 0.159 & 0.179 & 0.150 \\
L8 & 0.141 & 0.141 & 0.155 & 0.142 & 0.173 & 0.145 \\
L9 & 0.171 & 0.171 & 0.186 & 0.175 & 0.199 & 0.171 \\
L10 & 0.323 & 0.314 & 0.326 & 0.327 & 0.325 & 0.324 \\
L11 & 0.288 & 0.284 & 0.291 & 0.290 & 0.291 & 0.289 \\
L12 & 0.272 & 0.273 & 0.274 & 0.273 & 0.274 & 0.272 \\
\midrule
Mean & 0.164 & 0.163 & 0.178 & 0.169 & 0.194 & 0.167 \\
\bottomrule
\end{tabular}

\end{table}

\begin{table}[h!]
\centering
\caption{Ablation study - Favorita}
\label{tab:ablation-Favorita}
\begin{tabular}{lcccccc}
\toprule
Level & default & indep. copula & no reconc. & small subhier. & no xreg & arima \\
\midrule
L1 & 0.112 & 0.117 & 0.143 & 0.115 & 0.113 & 0.119 \\
L2 & 0.143 & 0.145 & 0.168 & 0.144 & 0.144 & 0.148 \\
L3 & 0.146 & 0.150 & 0.169 & 0.151 & 0.148 & 0.153 \\
L4 & 0.167 & 0.173 & 0.185 & 0.170 & 0.166 & 0.174 \\
L5 & 0.124 & 0.131 & 0.139 & 0.128 & 0.128 & 0.128 \\
L6 & 0.167 & 0.176 & 0.178 & 0.169 & 0.169 & 0.167 \\
L7 & 0.225 & 0.242 & 0.230 & 0.225 & 0.226 & 0.225 \\
L8 & 0.174 & 0.177 & 0.191 & 0.175 & 0.176 & 0.174 \\
L9 & 0.177 & 0.181 & 0.198 & 0.179 & 0.180 & 0.179 \\
L10 & 0.196 & 0.209 & 0.209 & 0.196 & 0.197 & 0.197 \\
L11 & 0.202 & 0.213 & 0.213 & 0.204 & 0.203 & 0.207 \\
L12 & 0.203 & 0.215 & 0.215 & 0.205 & 0.205 & 0.209 \\
L13 & 0.218 & 0.235 & 0.224 & 0.219 & 0.219 & 0.220 \\
L14 & 0.222 & 0.234 & 0.225 & 0.223 & 0.223 & 0.224 \\
L15 & 0.225 & 0.236 & 0.228 & 0.226 & 0.225 & 0.227 \\
L16 & 0.231 & 0.236 & 0.231 & 0.231 & 0.231 & 0.231 \\
\midrule
Mean & 0.183 & 0.192 & 0.197 & 0.185 & 0.185 & 0.186 \\
\bottomrule
\end{tabular}

\end{table}

\end{document}